\documentclass[%
reprint,
superscriptaddress,
amsmath,amssymb,
prb,
]{revtex4-2}
\usepackage{amsmath,amssymb,amsfonts,amsthm,mathtools}
\usepackage{pgfplots,pgfplotstable,physics,epstopdf}
\usepackage{amsthm}
\usepackage{CJK}
\usepackage{bm}
\usepackage{graphicx}
\usepackage{subfigure}
\usetikzlibrary{shadings}
\usepackage[colorlinks=true,linkcolor=blue,anchorcolor=red,citecolor=blue,urlcolor=blue]{hyperref}
\usepackage{appendix}
\usepackage{tikz,pgf}
\usepackage{pdfpages}
\usepackage{verbatim}

\makeatletter
\AtBeginDocument{\let\LS@rot\@undefined}
\makeatother

\def \K {{\mathcal{K}}}

\def \k {\bm{k}}

\def \H {\mathcal{H}}

\begin{document}

\title{Asymmetric real topology of conduction and valence bands}

\author{J. X. Dai}
\affiliation{Department of Physics and HK Institute of Quantum Science \& Technology, The University of Hong Kong, Pokfulam Road, Hong Kong, China}

\author{Chen Zhang}
\affiliation{National Laboratory of Solid State Microstructures and Department of Physics, Nanjing University, Nanjing 210093, China}

\affiliation{The University of HongKong Shenzhen Institute of Research and Innovation, Shenzhen 518057, China}

\author{Y. X. Zhao}
\email[]{yuxinphy@hku.hk}
\affiliation{Department of Physics and HK Institute of Quantum Science \& Technology, The University of Hong Kong, Pokfulam Road, Hong Kong, China}

\begin{abstract}
	Previously, it was believed that conduction and valence bands exhibit a symmetry: They possess opposite topological invariants (e.g., the Chern numbers of conduction and valence bands for the Chern insulator are $\pm C$). However, we present a counterexample: The second Stiefel-Whitney numbers for conduction and valence bands over the Klein bottle may be asymmetric, with one being nontrivial while the other trivial. Here, the Stiefel-Whitney classes are the characteristic classes for real Bloch functions under $PT$ symmetry with $(PT)^2=1$, and the Klein bottle is the momentum-space unit under the projective anticommutation relation of the mirror reflection reversing $x$ and the translation along the $y$ direction. The asymmetry originates from the algebraic difference of real cohomology classes over the Klein bottle and torus.  This discovery is rooted in the foundation of topological band theory, and has the potential to fundamentally refresh our current understanding of topological phases.
\end{abstract}
\maketitle

\section{Introduction}
Topological band theory originated from the TKNN invariant and Haldane model~\cite{TKNN,Haldanemodel}, and has been further developed through studying various topological insulators, superconductors, and semimetals~\cite{KaneRMP2010,SCZHANGRMP2011,SchnyderRMP2016,VishwanathRMP2018}. The basic concept is that with an energy gap, the topological configurations of the valence bands or the conduction bands can be characterized by symmetry-preserving topological invariants.

As far as we know, there is a general rule that governs the current theory: The conduction and valence bands have opposite values $N^{\pm}$ for a given topological invariant $N$, namely
\begin{equation}
	N^+=-N^-.
\end{equation}
This principle is referred to as the topological symmetry between the valence and conduction bands. There are numerous examples of this symmetry in action. For instance, in a Chern insulator, if the valence bands possess a first Chern number $C$, then the conduction bands will have $-C$. For the SSH model with the sublattice symmetry, the valence and conduction bands will respectively have winding numbers $\pm W$. For time-reversal invariant topological insulators in both two and three dimensions, the valence and conduction bands will have either simultaneously trivial or nontrivial $\mathbb{Z}_2$ topological invariants. More generally, all topological insulators or superconductors in the tenfold classifications adhere to this topological symmetry~\cite{SchnyderPRB2008,KitaevAIP2009}.

This article presents a counterexample to the topological symmetry between valence and conduction bands. Specifically, we show that the second Stiefel-Whitney number can be asymmetric over the Brillouin Klein bottle, i.e.,
\begin{equation}\label{eq:asymmetry}
	w_2^+\ne - w_2^-,
\end{equation}
where the Stiefel-Whitney (SW) numbers $w_{1,2}^{\pm}$ are valued in $ \mathbb{Z}_2=\{\bar{0},\bar{1}\}$.
When $w_2^-$  for the valence bands is nontrivial, $w_2^+$ for the conduction bands can be trivial.
The extraordinary result comes from the interplay between two recent research focuses, real topology and Brillouin Klein bottle. Asymmetric valence and conduction bands have been studied for unstable topological invariants like the Euler number in various systems~\cite{Bouhon2020PRB,bouhon2022arxiv}. This work focuses on the asymmetry for stable topological invariants.

Spacetime inversion symmetry of spinless particles implies that Bloch wavefunctions in the Brillouin zone are essentially real~\cite{YXZHAOPRL2016,YXZHAOPRL2017}. The corresponding characteristic classes for such systems are the SW classes, analogous to how Chern classes describe complex Bloch wavefunctions~\cite{YXZHAOPRL2017,BJYPRL2018,BJYPRX2019}. Recently, there has been significant research into $1$D and $2$D topological insulators and 3D nodal-line semimetals, which exhibit nontrivial first and second SW classes~\cite{BJYPRL2018,Park2019PRL,BJYPRX2019,ShengXLPRL2019,lee2020npj,WangPRL2020,ShengXLPRL2022,xue2023stiefel}. 

The Brillouin torus exhibits symmetric SW classes for valence and conduction bands. However, asymmetry arises, when reducing the momentum-space unit from the Brillouin torus to the Brillouin Klein bottle by a momentum-space glide reflection due to projective symmetry~\cite{ChenZYNC2022}. Since momentum-space nonsymmorphic symmetry was first introduced by~Ref.~\cite{ChenZYNC2022}, its general theory has been formulated within the framework of projective representations of crystallographic groups~\cite{Zhang2023PRL}. It has emerged as a prominent research focus, with intensive investigations in both artificial crystals~\cite{Li2023PRB, wang2023chess, ZHU2024SciBu, Lai2024PRA, Fonseca2024PRL, Hu2024PRL, Tao2024PRB, qiu2024octupole, konig2025exceptional, vaidya2025quantized} and condensed matter systems~\cite{Li2023PRB,Xiao2024PRX,cualuguaru2024arxiv}. Interestingly, the 2D Klein bottle has been generalized to the ten platycosms in momentum space~\cite{zhang2025brillouin}.

Over the Brillouin Klein bottle, the first SW class remains symmetric ($w_1^{-}=-w_1^{+}$) for valence and conduction bands. But, with certain nontrivial $w_1$, the second SW class becomes asymmetric [Eq.~\eqref{eq:asymmetry}].  Notably, this topological asymmetry originates from structural differences of cohomology rings for the Klein bottle and torus~\cite{Hatcherbook}, as will be elucidated. 

Our theory is explicitly demonstrated by constructing a 2D tight-binding model, which can catch immediate experimental interests. The asymmetry of the second SW class between valence and conduction bands results in distinct corner state patterns, compared to ordinary SW topological insulators~\cite{WangPRL2020}. 

Our work refreshes our comprehension of topological band theory at a fundamental level, and therefore can readily attract much experimental interest as well as pave the way for  exploring topological phenomena beyond the current framework. 
	
\section{Main results}
   Before diving into the technical details of proving the topological asymmetry, we first introduce our main result in concrete terms and then demonstrate it by a theoretical model.
   
   Under the projective symmetry algebra:
   \begin{equation}
   	\{M_x,L_y\}=0,
   \end{equation}
   the representation of $M_x$ acts in momentum space as the glide reflection $\mathcal{G}_x$ with~\cite{ChenZYNC2022}
   \begin{equation}\label{eq:m_glide}
   	\mathcal{G}_x:~(k_x,k_y)\mapsto (-k_x,k_y+\pi).
   \end{equation}
   This can be immediately derived as follows. The algebra can be recast as $M_xL_yM_x^{-1}=-L_y$. In momentum space, $L_y=e^{ik_y}$, and therefore $M_xe^{ik_y}M_x^{-1}=e^{i(k_y+\pi)}$. Thus, apart from inversing $k_x$, $M_x$ translates $k_y$ by $\pi$, as in Eq.~\eqref{eq:m_glide}.
   
   For a non-interacting tight-binding model $\mathcal{H}(\bm{k})$, the operator of $M_x$ may be written as $M_x=U_M\mathcal{G}_x$, and the commutation relation $[\mathcal{H}(\bm{k}),M_x]=0$ can be expanded as
   \begin{equation}\label{eq:Gxsymmetry}
   	U_M\mathcal{H}(\bm{k})U_M^\dagger=\mathcal{H}(-k_x,k_y+\pi).
   \end{equation}
   The mirror reflection $M_x$ is represented as the momentum-space glide reflection symmetry $\mathcal{G}_x$, which not only inverts $k_x$ but translates $k_y$ by  $\pi$.  
   
\begin{figure}
   	\centering
   	\includegraphics[width=\columnwidth]{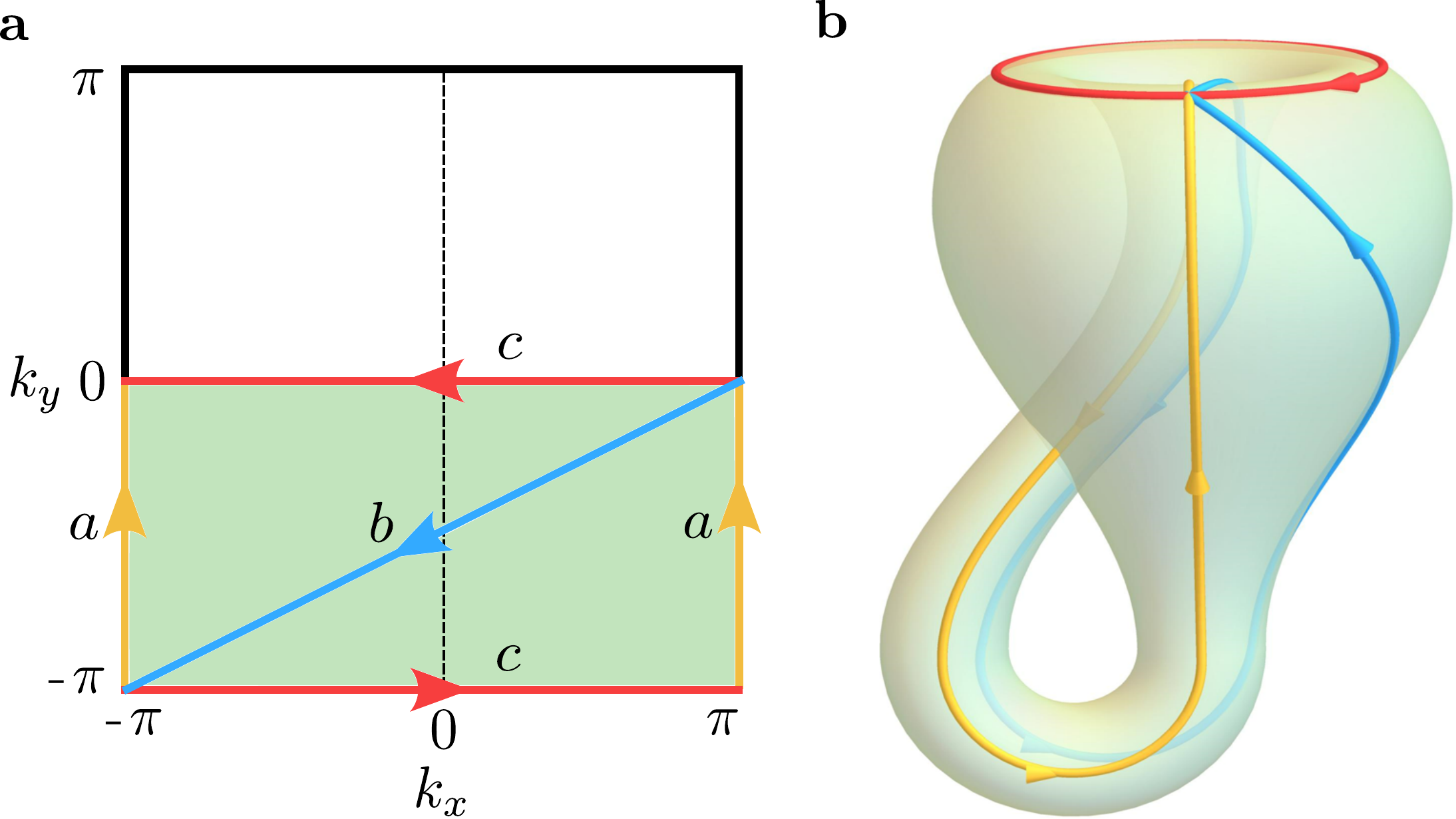}
   	\caption{The Brillouin Klein bottle. \textbf{a} The fundamental domain and the boundary identification relations under the free action of the glide reflection. \textbf{b} The resulted Klein bottle under the boundary identification relations indicated in (\textbf{a}). \label{fig:Glide-Mirror}}
   \end{figure}

   As illustrated in Fig.~\ref{fig:Glide-Mirror}, $\mathcal{G}_x$ acts freely on momentum space, and induces the anti-periodic boundary conditions between the $k_x$ subsystems with $k_y=-\pi$ and $k_y=0$.  This leads to the Klein bottle as the momentum-space unit in the presence of $\mathcal{G}_x$.
   
   To enable essentially real bands, we need the spacetime inversion symmetry $PT$ with $(PT)^2=1$. Without loss of generality, we can always adopt the convention $PT=\mathcal{K}$ with $\mathcal{K}$ the complex conjugation by locally choosing an appropriate basis in the Brillouin zone~\cite{YXZHAOPRL2017}. The band topology in one and two dimensions can be characterized by Stiefel-Whitney classes~\cite{YXZHAOPRL2017,BJYPRL2018,BJYPRX2019,Milnorcharacteristic1974}. Below, we assume an energy gap and introduce the first and second Stiefel Whitney classes $w_{1,2}^{-}$ of valence bands, while $w_{1,2}^{+}$ are defined completely in parallel for conduction bands.
   
   For valence bands of an insulator, the integration of $w_1^-$ over a $1$D closed path $C$ in the Brillouin zone just corresponds to the Berry phase $\nu_1^-(C)$ over $C$ modulo $2\pi$, recalling that $\nu_1^-(C)$ is quantized into multiples of $\pi$ by $PT$ symmetry.
   
   Over the Klein bottle $K$, $w_2^-$ can be completely determined by its integration $w_2^-(K)$. In more concrete terms, it can be characterized by the topological (homotopy) class of the transition function of valence bands over the large circle $a$.  With $N$ real valence bands, the topological classification is given by $\pi_1[\mathrm O(N)]=\mathbb{Z}_2$ for $N\ge 3$, i.e., $w_2^-$ is nontrivial if and only if the transition function, as a mapping from $a$ to $\mathrm O(N)$, belongs to the nontrivial class of $\pi_1[\mathrm O(N)]=\mathbb{Z}_2$. For $N=2$, $\pi_1[\mathrm O(2)]=\mathbb{Z}$, and the transition function is classified by the winding number $W\in \mathbb{Z}$. $w_2^-$ is nontrivial if and only if $W$ is odd. This is because if trivial bands are added to mix with the two bands, the transition functions with odd (even) winding numbers are nontrivial (trivial).
   
   We are now ready to concisely state the main result of our work: The asymmetry of $w_2^{\pm}$ for valence and conduction bands exists if and only if
   \begin{equation}\label{eq:c_phase}
   	\nu^-_1(c)=\pi \mod 2\pi.
   \end{equation}
   In other words, to have the topological asymmetry, we only need to form a nontrivial Berry phase over the edge $c$ with the anti-periodic boundary conditions.

   \section{Model illustration} 
  
   Before diving into the formal proof of the asymmetry, let us first illustrate our result by constructing a concrete model. 
   
  To construct such a model, first the model should preserve  both $PT$ symmetry and the mirror symmetry $M_x$ with $\{M_x,L_y\}=0$ so that the momentum-space unit is the Klein bottle.  Then, we assume it to be a three-band model with two valence bands and a conduction band. Accordingly, the second Stiefel-Whitney class of the conduction band is automatically trivial, and we only need to focus on the valence bands. Finally, we just need to adjust the parameters such that each one-dimensional $k_y$-subsystem possesses nontrivial valence-band Berry phase, which guarantees the nontrivial second Stiefel-Whitney class for the valence bands according to Eq.~\eqref{eq:c_phase}.
   
   Following the above general strategy, the model we figure out is given by
   \begin{equation}\label{eq:model}
   	\H(\k)=\begin{bmatrix}
   		q_0(\k) & q_1(\k) &q_2(\k)\\
   		q_1(\k) &q_3^+(\k)&q_4(\k)\\
   		q_2(\k)&q_4(\k)&q_3^-(\k)
   	\end{bmatrix},
   \end{equation} 
   where $q_0(\k)=\cos k_x$, $q_1(\k)=\sin k_x\cos k_y$, $q_2(\k)=\sin k_x\sin k_y$, $q_3^\pm(\k)=(1-\cos k_x)(1\pm\cos 2k_y)/2-1$, and $q_4(\k)=(1-\cos k_x)\sin k_y\cos k_y$. It may be more illuminating to rewrite it as
   \begin{equation}\label{eq:diagonalization}
   	\H(\k)=\mathcal{O}(\k)\begin{bmatrix}
   		1&&\\&-1&\\&&-1\end{bmatrix}\mathcal{O}^T(\k),
   \end{equation}
   with
   \begin{equation}\label{eq:Orthomatrix}
   	\mathcal{O}(\k)=\exp \left[\frac{k_x}{2}(\cos k_y L_z-\sin k_y L_y)\right].
   \end{equation}
   Here, $L_i$ with $i=x,y,z$ are the standard basis of the Lie algebra of $\mathrm{SO}(3)$ with $L_i$ generating the rotation along the $i$ axis. The concrete representations of $L_i$'s can be found in Appendix~\ref{A}.

   The symmetry operators are given by $PT=\mathcal K$ and $M_x=\mathcal G_x$. $PT=\mathcal K$ is preserved since $L_i$ are skew symmetric real matrices, and the preservation of $M_x=\mathcal G_x$ [Eq.~\eqref{eq:Gxsymmetry}] can be seen from $\mathcal{O}(-k_x,k_y+\pi)=\mathcal{O}(k_x,k_y)$.
   
   We now calculate the Berry phase $\nu_1^{\pm}(c)$ along the closed path $c$, which is parameterized by $(k_x,0)$ with $k_x\in [-\pi,\pi)$. It is easy to derive the eigenstates along $c$ explicitly as
   $|\psi^+(k_x,0)\rangle=e^{ik_x/2}(\cos k_x/2, \sin k_x/2, 0)^T$ for the conduction band, and  $|\psi^-_1(k_x,0)\rangle=e^{ik_x/2}(-\sin k_x/2, \cos k_x/2, 0)^T$, $|\psi^-_2(k_x,0)\rangle=(0,0,1)^T$ for the two valence bands. Note that all the eigenstates are periodic along $k_x$. Then, the Berry phases can be directly calculated as $\nu_1^{\pm}(c)=\pi \mod 2\pi$, and therefore Eq.~\eqref{eq:c_phase} is satisfied. For completeness, we note that $\nu_1^{\pm}(a)=\pi \mod 2\pi$ and $\nu_1^{\pm}(b)=0 \mod 2\pi$, with $\nu_1^{\pm}(a)+\nu_1^{\pm}(b)+\nu_1^{\pm}(c)=0\mod 2\pi$.
   
   Hence, according to Eq.~\eqref{eq:c_phase}, the model should have the asymmetry, $w_2^+\ne w_2^-$, which is verified in the following. Since there is only one conduction band, $w_2^+$ is automatically trivial with $\nu^+_2(K)=0$. Hence, we only need to focus on the two valence bands, for which the eigenstates are given by
   \begin{equation}\label{eq:valence-bands}
   	|\psi^-_\alpha(\k)\rangle=\mathcal{O}(\k)|d_\alpha\rangle
   \end{equation}
   with $|d_1\rangle =(0,1,0)^T$ and $|d_2\rangle =(0,0,1)^T$. We should look into the transition function $t(k_y)$ gluing $a$ at $k_x=\pi$ to the same $a$ at $k_x=-\pi$, which is given by
   \begin{equation}\label{eq:transition}
   	|\psi_\beta^-(-\pi,k_y)\rangle =\sum_\alpha |\psi_{\alpha}^-(\pi,k_y)\rangle [t(k_y)]_{\alpha\beta}.
   \end{equation}
   From Eqs.\eqref{eq:valence-bands} and \eqref{eq:transition}, the transition function can be derived as 
   \begin{equation}
   	t(k_y)=\begin{bmatrix}
   		\cos 2k_y & -\sin 2k_y\\
   		\sin 2k_y & \cos 2k_y
   	\end{bmatrix}\begin{bmatrix}
   		-1 & 0\\
   		0 &  1
   	\end{bmatrix}.
   \end{equation}
   Then, the transition function, as a mapping from the closed path $a$ to $O(2)$, has the winding number,
   \begin{equation}\label{eq:widing}
   	W[t]=-\frac{1}{4\pi}\int_{0}^{\pi} dk_y~\mathrm{Tr}J t(k_y)\partial_{k_y}[t(k_y)]^{-1}=1.
   \end{equation}
   Here, $J=-i\sigma_2$ with $\sigma_2$ the second Pauli matrix.
   According to our earlier introduction to the SW classes, $w_2^-$ is nontrivial. Thus, we have verified $w_2^+\ne w_2^-$. The boundary states can be found in Appendix~\ref{B}.
	
  \section{Asymmetric SW classes}
After giving the example model, we proceed to formally prove the asymmetry over the Brillouin Klein bottle. To be self-contained, a brief introduction to the algebraic theory of cohomology is given in the Appendix~\ref{C}. The proof is based on the algebraic structure of SW classes for valence and conduction bands, as introduced in this section. 

For each $\bm{k}$, the eigenstates above (below) the gap span the conduction/valence space $E^{\pm}(\bm{k})$, both of which are real vector spaces. Over the momentum-space unit $\mathcal{M}$, we hence obtain continuous distributions $E^{\pm}_\mathcal{M}$ of conduction and valence spaces, respectively. Here, $\mathcal{M}$ is the Brillouin torus $T$ or Klein bottle $K$. The total SW classes $w^\pm=w(E^{\pm}_\mathcal{M})\in H^*(X,\mathbb{Z}_2)$ for conduction and valence bands can be concisely expressed as~\cite{Milnorcharacteristic1974}
\begin{equation}\label{eq:total_classes}
	w^\pm =1+w^{\pm}_1+w^{\pm}_2.
\end{equation}

For each momentum $\bm{k}\in \mathcal{M}$, the total Hilbert space $V_\mathcal{M}(\bm{k})$ is also real, and is spanned by the valence and conduction states, i.e., $V_\mathcal{M}(\bm{k})=E^+_\mathcal{M}(\bm{k})\oplus E^-_\mathcal{M}(\bm{k})$, and therefore we may write $V_\mathcal{M}=E^+_\mathcal{M}\oplus E^-_\mathcal{M}$.  A significant observation is that $V_\mathcal{M}$ is always a flat distribution of vector spaces, and therefore is topologically trivial. This means $w(V_\mathcal{M})=1$, or alternatively, $w_n(V_\mathcal{M})=0$ for $n>0$. According to the general theory of characteristic classes, $w(E^+_\mathcal{M}\oplus E^-_\mathcal{M})=w(E^+_\mathcal{M})w(E^-_\mathcal{M})$~\cite{Milnorcharacteristic1974}.  Consequently, 
\begin{equation}\label{eq:triviality}
	w^+w^-=1.
\end{equation}

Equations \eqref{eq:total_classes} and \eqref{eq:triviality} lead to $1+w^+_1+w^+_2={1}/({1+w^-_1+w^-_2})=1+(w^-_1+w^-_2)+(w^-_1+w^-_2)^2=1+w^-_1+(w^-_2+w^-_1w^-_1)$. It is noteworthy that $-1=1 \mod 2$, and terms with order higher than $2$ vanish in the expansion since $\mathcal{M}$ is two dimensional. Hence, we have the relations of first and second SW classes for valence and conduction bands:
\begin{align}
	w^+_1 &= w^-_1, \label{eq:1st_relation}\\
	w^+_2 &=w^-_2+w^-_1w^-_1 \label{eq:2nd_relation}.
\end{align}
The form of the second relation \eqref{eq:2nd_relation} is manifestly asymmetric between valence and conduction bands. As we shall show, over the Brillouin torus $T$, the symmetry is in fact preserved, because we always have $w^-_1w^-_1=0$ over $T$. However, the symmetry is broken over the Brillouin Klein bottle $K$, where $w^-_1w^-_1$ may be nonzero. 

\begin{figure}
	\centering
	\includegraphics[width=\columnwidth]{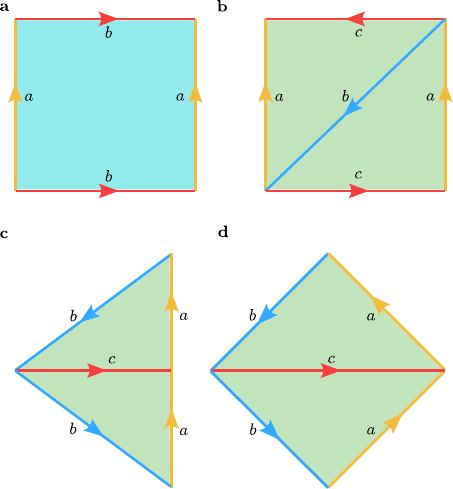}
	\caption{Brillouin torus and Klein bottle. \textbf{a} The boundary identification relations of the torus. \textbf{b}-\textbf{d} The boundary identification relations of the Klein bottle. In (\textbf{c}), the square is cut through $b$ and re-glued along $c$. (\textbf{c}) is deformed to be (\textbf{d}). \label{fig:2D_model}}
\end{figure}

\textit{Symmetry over the Brillouin torus.} On the Brillouin torus $T$, there are two independent $1$D nonzero cycles $a$ and $b$ as illustrated in Fig.~\ref{fig:2D_model}(a). The linear space $H_1(T,\mathbb{Z}_2)$ is spanned by $a$ and $b$, i.e., $H_1(T,\mathbb{Z}_2)=\langle a, b\rangle_{\mathbb{Z}_2}$, consisting of four linear combinations $n_aa+n_bb$ with $n_{a,b}\in\mathbb{Z}_2$. Then, we have the dual linear space $H^1(T,\mathbb{Z}_2)=\langle \alpha, \beta \rangle_{\mathbb{Z}_2}$, where $\alpha$ and $\beta$ are dual to $a$ and $b$, respectively, satisfying
\begin{equation}\label{eq:dual_basis}
	\begin{split}
		\alpha(a)=1,\quad \alpha(b)=0,\\
		\beta(a)=0,\quad \beta(b)=1.
	\end{split}
\end{equation}
Then, $H^1(T,\mathbb{Z}_2)$ consists of $m_\alpha\alpha+m_\beta\beta$ with $m_{\alpha,\beta}\in\mathbb{Z}_2$. Note that $(m_\alpha\alpha+m_\beta\beta)(n_a a+n_b b)=m_\alpha n_a+m_\beta n_b$.

There is only one nontrivial $2$D cycle, namely the whole torus $T$, and therefore one nontrivial cocycle $\tau$ dual to $T$ with $\tau(T)=1$. Formally, $H_{2}(T,\mathbb{Z}_2)=\langle T\rangle_{\mathbb{Z}_2}$ and $H^{2}(T,\mathbb{Z}_2)=\langle \tau\rangle_{\mathbb{Z}_2}$, where $(m_\tau\tau)(n_TT)=m_\tau n_T$ with $n_T,m_\tau\in\mathbb{Z}_2$. The product of two $1$D cocycles is a $2$D cocycle, and for torus the multiplication in $H^1(T,\mathbb{Z}_2)$ is given by~\cite{Hatcherbook}
\begin{equation}\label{eq:torus_wedge}
	\alpha\alpha=\beta\beta=0,\quad \tau=\alpha\beta.
\end{equation}
The product of two generic $1$D terms is defined by requiring linearity.

We are now ready to explain why the second SW class is still symmetric for conduction and valence bands over the Brillouin torus. The first SW class $w^-_1$ of the valence bands is determined by the Berry phases $\nu^{1}_a$ and $\nu^1_b$ over cycles $a$ and $b$ on $T$, respectively. In the linear combination $w^-_1=m_\alpha \alpha+m_\beta \beta$, $m_{\alpha}=\nu^1_a/\pi\mod 2$ and $m_{\beta}=\nu^1_b/\pi\mod 2$. It is significant to observe that there are only three possibilities for $w^-_1$, i.e., $w^-_1=\alpha,\beta$ or $\alpha+\beta$, corresponding to $\nu^1$ being nontrivial over $a, b$ and $c=a+b$, respectively.  Clearly, from \eqref{eq:torus_wedge}, $w^-_1w^-_1$ always vanishes in all three cases. Hence, despite the apparent asymmetry in \eqref{eq:2nd_relation}, the second SW class is in fact symmetric for conduction and valence bands.

\textit{Asymmetry over the Brillouin Klein bottle.}
With coefficients $\mathbb{Z}_2$, the linear spaces of homology and cohomology of the Klein bottle $K$ are isomorphic to those of the torus. We choose $a$ and $b$ in Fig.\ref{fig:2D_model}(b) as base vectors for $H_1(K,\mathbb{Z}_2)$, namely $H_1(K,\mathbb{Z}_2)=\langle a, b\rangle_{\mathbb{Z}_2}$. It is noteworthy that the two cycles $a$ and $b$ are topologically on an equal footing. To see this, we first cut through $b$ in Fig.\ref{fig:2D_model}(b), and then identify two $c$ edges, which gives Fig.\ref{fig:2D_model}(c). We can continuously deform Fig.\ref{fig:2D_model}(c) to Fig.\ref{fig:2D_model}(d), where $a$ and $b$ are clearly on the equal footing. Still, $H^1(K,\mathbb{Z}_2)=\langle \alpha,\beta \rangle_{\mathbb{Z}_2}$ with $\alpha$ and $\beta$ the dual basis as in \eqref{eq:dual_basis}. $H_2(K,\mathbb{Z}_2)=\langle K \rangle_{\mathbb{Z}_2}$ and $H^2(K,\mathbb{Z}_2)=\langle \kappa \rangle_{\mathbb{Z}_2}$ with $\kappa(K)=1$. 

The topological difference of the Klein bottle from the torus is embodied in the multiplication in $H^*(K,\mathbb{Z}_2)$, which is given by~\cite{Hatcherbook}
\begin{equation}\label{eq:K_wedge}
	\alpha\alpha=\beta\beta=\kappa, \quad \alpha\beta=0.
\end{equation}
One may compare \eqref{eq:K_wedge} with \eqref{eq:torus_wedge}. Notably, they are the only two algebras over the linear space $\mathbb{Z}_2\oplus\mathbb{Z}_2$ that are symmetric in the base vectors.

Now, over the Klein bottle, asymmetry can arise from the multiplications \eqref{eq:K_wedge}. If the Berry phases of valence bands along $a$ and $b$ satisfy $(\nu^1(a),\nu^1(b))=(\pi,0)$ or $(\nu^1(a),\nu^1(b))=(0,\pi)$, the first SW classes are given by 
\begin{equation}\label{eq:w1_K}
	w_1^{\pm}=\alpha~\mathrm{or}~\beta.
\end{equation}
Then, according to \eqref{eq:K_wedge}, $w^-_1w^-_1=\kappa$. Equation \eqref{eq:2nd_relation} further implies two possibilities:
\begin{equation}
	(w^+_2, w^-_2)=(\kappa,0)~\mathrm{or}~ (0,\kappa),
\end{equation}
both of which are asymmetric for conduction and valence bands. Furthermore, if the Berry phases over both $a$ and $b$ are nontrivial, then $w^-_1=\alpha+\beta$, and the second SW class is still symmetric, since $(\alpha+\beta)^2=\kappa+\kappa=0$. 

The above arguments lead to our main result, namely that the necessary and sufficient condition for the asymmetry is Eq.~\eqref{eq:w1_K}. Moreover, Eq.~\eqref{eq:w1_K} is equivalent to Eq.~\eqref{eq:c_phase}, as explained in the following.  $a$, $b$ and $c$ together enclose a triangle [Fig.~\ref{fig:2D_model}(b-d)]. But because of $PT$ symmetry, the Berry curvature is zero everywhere in the Brillouin zone, and therefore $\nu^1(c)=\nu^1(a)+\nu^1(b) \mod 2\pi$. Thus, Eq.~\eqref{eq:c_phase} is satisfied, if and only if one and only one of $\nu^1(a)$ and $\nu^1(b)$ is equal to $\pi\mod 2\pi$, i.e., Eq.~\eqref{eq:w1_K} holds.

\section{Summary and discussions}
In summary, we have established that the characteristic class of the valence bands may not be opposite to that of the conduction bands by both rigorous theoretical proof and example demonstration.  This is an extraordinary discovery in the foundation of topological band theory that may disprove the common wisdom held by the community.

Recently, various metamaterials have proved their incredible power to simulate complex lattice models with nontrivial topological properties~\cite{LuNaturePhotonics2014,YangPRL2015,HuberNaturePhysics2016,Zhang2018Adip,Ronny_2018np,Peterson_2018nature,Serra_Garcia_2018nature,OzawaRMP2019,MaGC_2019nature,Yu_Zhao_NSR2020,NiNatureCommu2020}. Especially, the Brillouin Klein bottle and real topologies under $PT$ symmetry are two recent hot topics under intensive experimental investigation~\cite{Pu2023PRB,Li2023PRB,ZHU2024SciBu,Fonseca2024PRL,Hu2024PRL,Tao2024PRB,xue2023stiefel,Tomas2019Science,YangEr2020PRL,guo2021nature,jiang2021NC,jiang2024SB}.  It is expected that our model can be realized  and our theory can be confirmed by artificial crystals made of various metamaterials.

Finally, let us comment on the boundary states of the topological insulator with asymmetric valence-conductance band topologies. The asymmetry of $w_2^\pm $ exists only when $w_1^\pm$ is nontrivial with $\nu^\pm (c)=\pi \mod 2\pi$. The nontrivial Berry phases along $1$D $k_x$-subsystems mean the insulator has nontrivial weak topology, leading to a nearly flat in-gap band on the $y$ edges. Considering a rectangular geometry, the real Klein bottle insulator with nontrivial $w_2^\pm$ will possess corner states at all four corners. In contrast, an ordinary real topological insulator with nontrivial $w_2^\pm$ generically hosts corner states only at a pair of inversion-related corners, rather than at all corners~\cite{WangPRL2020}. This difference can be attributed to the mirror symmetry $M_x$, which together with the anti-commutation relation $\{M_x,L_y\} =0$ reduces the momentum-space unit from Brillouin torus to the Brillouin Klein bottle. The mirror symmetry relates a pair of inversion-related corners to the other pair of inversion-related corners in a rectangular geometry. Thus, once such a Klein-bottle insulator has the nontrivial second Stiefel-Whitney class, all corners host mid-gap states.

\begin{acknowledgements}
This work was supported by the GRF of Hong Kong (No. 17301224), and the NSFC/RGC JRS grant (No. 12161160315).
\end{acknowledgements}
\appendix

\section{The topology of the model in the manintext}\label{A}
The Hamiltonian in the maintext can be written as
\begin{equation}\label{eq:Hamiltonian}
	\H(\k)=\mathcal{O}(\k)\begin{bmatrix}
		1&&\\&-1&\\&&-1\end{bmatrix}\mathcal{O}^T(\k),
\end{equation}
with
\begin{equation}\label{eq:Orthomatrix2}
	\begin{split}
		\mathcal{O}(\k)&=\exp \left[\frac{k_x}{2}(\cos k_y L_z-\sin k_y L_y)\right].
	\end{split}
\end{equation}
Here, $L_i$ with $i=x,y,z$ are given by
\begin{equation*}
	L_x=\begin{bmatrix}
		0&0&0\\0&0&-1\\0&1&0
	\end{bmatrix},~	L_y=\begin{bmatrix}
		0&0&1\\0&0&0\\-1&0&0
	\end{bmatrix},~L_z=\begin{bmatrix}
		0&-1&0\\1&0&0\\0&0&0
	\end{bmatrix}.
\end{equation*}
They satisfy the commutation relations
\begin{equation*}
	[L_x,L_y]=L_z,~[L_z,L_x]=L_y,~[L_y,L_z]=L_x.
\end{equation*}
Meanwhile, the explicit form of $\mathcal O(\k)$ in Eq.~\eqref{eq:Orthomatrix2} is given by
\begin{equation}\label{eq:explicitO}
	\mathcal O(\k)=\begin{bmatrix}
		p_0(\k) & -p_1(\k) &-p_2(\k)\\
		p_1(\k) &p_3^+(\k)&p_4(\k)\\
		p_2(\k)&p_4(\k)&p_3^-(\k)
	\end{bmatrix},
\end{equation} 
where $p_0(\k)=\cos \frac{k_x}{2}$, $p_1(\k)=\sin \frac{k_x}{2}\cos k_y$, $p_2(\k)=\sin \frac{k_x}{2}\sin k_y$, $p_3^\pm(\k)=(
\cos \frac{k_x}{2}-1)(1\pm\cos 2k_y)/2+1$, and $p_4(\k)=(\cos \frac{k_x}{2}-1)\sin k_y\cos k_y$.

From Eqs.~\eqref{eq:Hamiltonian} and \eqref{eq:explicitO}, we can obtain the conduction eigenfunction $|\psi^+(\k)\rangle$ and valence eigenfunctions  $|\psi^-_{1,2}(\k)\rangle$ as.
\begin{equation}\label{eq:eigenstates}
	\begin{split}
		&|\psi^+(\k)\rangle=\mathcal O(\k)|d_0\rangle=(p_0,p_1,p_2)^T,\\
		&|\psi^-_1(\k)\rangle=\mathcal O(\k)|d_1\rangle=(-p_1,p_3^+,p_4)^T,\\
		&|\psi^-_2(\k)\rangle=\mathcal O(\k)|d_2\rangle=(-p_2,p_4,p_3^-)^T.
	\end{split}
\end{equation} 
Here, $|d_0\rangle=(1,0,0)^T$, $|d_1\rangle=(0,1,0)^T$ and $|d_2\rangle=(0,0,1)^T$. Then the Berry phases $\nu_1^\pm$ for conduction and valence bands can be calculated along cycles $a$, $b$ and $c$ in Fig.~\ref{fig:Glide-Mirror}(a) as
\begin{equation}\label{eq:Berry_phase}
	\begin{split}
		&\nu_1^+=\int_{S^1}dk_i~\mathrm i\langle \psi^+(k_i)|\partial_{k_i}|\psi^+(k_i)\rangle,\\
		&\nu_1^-=\int_{S^1}dk_i~\mathrm i\sum_{j=1}^2\langle \psi^-_j(k_i)|\partial_{k_i}|\psi^-_j(k_i)\rangle.
	\end{split}
\end{equation}
It is noteworthy that $k_x\in[-\pi,\pi)$ while $k_y\in[-\pi,0)$ in Eq.~\eqref{eq:Berry_phase}. 

Along the cycle $a$ with $k_x=\pi$ and $k_y\in[-\pi,0)$, the eigenstates are 
\begin{equation}\label{eq:eigien_a}
	\begin{split}
		&|\psi^+\rangle_a=e^{ik_y}(0,\cos k_y,\sin k_y)^T,\\
		&|\psi^-_1\rangle_a=e^{ik_y}(-\cos k_y,\sin^2 k_y,-\sin k_y\cos k_y)^T,\\
		&|\psi^-_2\rangle_a=(-\sin k_y,-\sin k_y\cos k_y,\cos^2 k_y)^T.
	\end{split}
\end{equation}
Here, the phases $e^{ik_y}$ are added to make the eigenfunctions well-defined on the cycle $a$. Substituting Eq.~\eqref{eq:eigien_a} into Eq.~\eqref{eq:Berry_phase}, the Berry phases can be directly calculated as $\nu_1^\pm(a)=\pi$.

Along the cycle $b$ with $k_x=2k_y+\pi$ and $k_y\in[-\pi,0)$, the eigenstates are 
\begin{equation}\label{eq:eigien_b}
	\begin{split}
		|\psi^+\rangle_b=(-&\sin k_y,\cos^2 k_y,\sin k_y\cos k_y)^T,\\
		|\psi^-_1\rangle_b=(-&\cos^2 k_y,-\sin k_y\cos^2 k_y+\sin^2 k_y,\\
		&-(\sin k_y+1)\sin k_y\cos k_y)^T,\\
		|\psi^-_2\rangle_b=(-&\sin k_y\cos k_y,-(\sin k_y+1)\sin k_y\cos k_y,\\
		&-\sin^3 k_y+\cos^2 k_y)^T.
	\end{split}
\end{equation}
Substituting Eq.~\eqref{eq:eigien_b} into Eq.~\eqref{eq:Berry_phase}, the Berry phases can be directly calculated as $\nu_1^\pm(b)=0$.

Along the cycle $c$ with $k_x\in[-\pi,\pi)$ and $k_y=0$, the eigenstates are
\begin{equation}\label{eq:eigien_c}
	\begin{split}
		&|\psi^+\rangle_c=e^{ik_x/2}(\cos k_x/2, \sin k_x/2, 0)^T,\\
		&|\psi^-_1\rangle_c=e^{ik_x/2}(-\sin k_x/2, \cos k_x/2, 0)^T,\\
		&|\psi^-_2\rangle_c=(0,0,1)^T.
	\end{split}
\end{equation}
Here, the phases $e^{ik_x/2}$ are added to make the eigenfunctions well-defined on the cycle $c$. Substituting Eq.~\eqref{eq:eigien_c} into Eq.~\eqref{eq:Berry_phase}, the Berry phases can be directly calculated as $\nu_1^\pm(c)=\pi$.
Clearly, the Berry phase for cycles $a$, $b$, and $c$ satisfy that
\begin{equation*}
	\nu_1(a)+\nu_1(b)+\nu_1(c)=0 \mod 2\pi.
\end{equation*}

Now, let us show the derivation details of the transition function $t(k_y)$ on the boundary $a$. $|\psi_{1,2}^-(\k)\rangle$ in Eq.~\eqref{eq:eigenstates} are not periodic along the $k_x$ direction, namely, $|\psi_{1,2}^-(k_x,k_y)\rangle\ne|\psi_{1,2}^-(k_x+2\pi,k_y)\rangle$. In Fig.~\ref{fig:Glide-Mirror}(a), the valence eigenfunctions are glued by the transition function $t(k_y)$ on the boundary $a$, which corresponds to both $k_x=-\pi$ and $k_x=\pi$. And the transition function is given by
\begin{equation}\label{eq:transition2}
	|\psi_\beta^-(-\pi,k_y)\rangle =\sum^2_{\alpha=1} |\psi_{\alpha}^-(\pi,k_y)\rangle [t(k_y)]_{\alpha\beta}.
\end{equation}
From Eq.~\eqref{eq:eigenstates}, we obtain the valence eigenstates $|\psi_{1,2}^-(\pm\pi,k_y)\rangle$ on the boundary $a$:
\begin{equation}\label{eq:psikya}
	\begin{split}
		|\psi_1^-(-\pi,k_y)\rangle=(\cos k_y,\sin^2 k_y,-\sin k_y\cos k_y)^T,\\
		|\psi_2^-(-\pi,k_y)\rangle=(\sin k_y,-\sin k_y\cos k_y,\cos^2 k_y)^T,\\
		|\psi_1^-(\pi,k_y)\rangle=(-\cos k_y,\sin^2 k_y,-\sin k_y\cos k_y)^T,\\
		|\psi_2^-(\pi,k_y)\rangle=(-\sin k_y,-\sin k_y\cos k_y,\cos^2 k_y)^T.
	\end{split}
\end{equation}
Substituting Eq.~\eqref{eq:psikya} into Eq.~\eqref{eq:transition2}, the transition function $t(k_y)$ is simply given by
\begin{equation}
	t(k_y)=\begin{bmatrix}
		\cos 2k_y & -\sin 2k_y\\
		\sin 2k_y & \cos 2k_y
	\end{bmatrix}\begin{bmatrix}
		-1 & 0\\
		0 &  1
	\end{bmatrix}.
\end{equation}
Then, by the topological invariant calculation equation \eqref{eq:widing}, we obtain the topological invariant $\nu_2^-=1$. \\

\section{The boundary states of the model in the maintext}\label{B}
This model possesses several topological invariants, leading to various topological boundary states under different boundary opening conditions.

\begin{figure}
	\centering
	\includegraphics[width=\columnwidth]{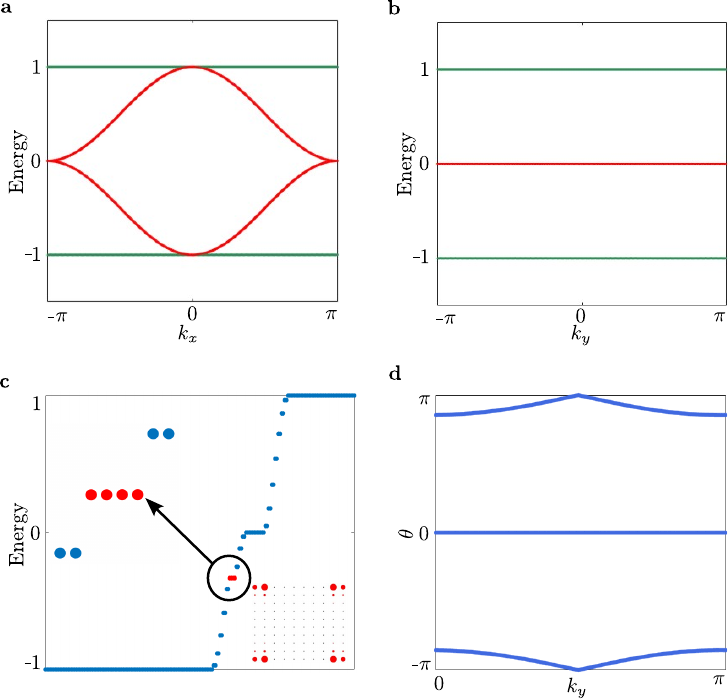}
	\caption{Boundary states under different boundary opening conditions. \textbf a and \textbf b show the edge states when the boundary is opened along the $x$ and $y$ directions, respectively. \textbf c Four corner states appear at the corners of a $10\times 10$ unit cell square sample. \textbf d The Wilson loop of $\H_t(\k)$ in Eq.~\eqref{eq:Ht} constructed by $\H(\k)$ in Eq.~\eqref{eq:model}.}
	\label{fige1}
\end{figure}

As shown in Fig. \ref{fige1}(a), helical edge states emerge when the boundary is opened along the $x$ direction. Consider the one-dimensional subsystems $\H_{k_x}(k_y)$ ($k_x$ is fixed) parametrized by $k_y\in[-\pi,0)$ in the Brillouin Klein bottle, only $\H_\pi(k_y)$ and $\H_0(k_y)$ are defined on the cycles.  Their Berry phases are $\pi$ and $0$, respectively. Correspondingly, $\H_\pi(k_y)$ possesses in-gap boundary states, while for $\H_0(k_y)$, the boundary states are mixed with the bulk states. Therefore, the helical edge states are protected by the topology asymmetry condition $\nu^1(a)\ne\nu^1(b)$.

In contrast, when the boundary is opened along the $y$ direction, the boundary states appear as flat bands in the energy spectrum, as illustrated in Fig. \ref{fige1}(b). This is because each one-dimensional subsystem $\H_{k_y}(k_x)$ with fixed $k_y$ has a nontrivial $w_1$. 

If we take a square sample where the periodic boundary condition is completely broken, we can find four corner states distributed at the four corners, as depicted in Fig. \ref{fige1}(c). These corner states are protected by the nontrivial $w_2^-$. 

In conclusion, the model \eqref{eq:Hamiltonian} is a first-order topological insulator with helical edge states, a weak topological insulator with flat edge states, and a second-order topological insulator with corner states.

\section{Cohomology and SW class}\label{C}
	Let us start with introducing some basics of (co)homology with $\mathbb{Z}_2$ coefficients. $\mathbb{Z}_2$ is the smallest number field consisting of only two elements $0$ and $1 \mod 2$. The $n$th homology group $H_n(X,\mathbb{Z}_2)$ is the linear space spanned by $n$D cycles with $\mathbb{Z}_2$ as the ground field. An $n$D cycle is an $n$D subspace without boundary, and if a set of cycles form the boundary of an $(n+1)$D subspace of $X$, the sum of them is defined to be zero. $H^n(X,\mathbb{Z}_2)$ consists of all linear functions from $H_n(X,\mathbb{Z}_2)$ to $\mathbb{Z}_2$; i.e., $H^n(X,\mathbb{Z}_2)$ is the dual linear space of $H_n(X,\mathbb{Z}_2)$. To consider all dimensions together, we introduce the total direct sums $H_{*}(X,\mathbb{Z}_2)=\oplus_{n} H_n(X,\mathbb{Z}_2)$ and $H^{*}(X,\mathbb{Z}_2)=\oplus_{n}H^n(X,\mathbb{Z}_2)$. Moreover, $H^{*}(X,\mathbb{Z}_2)$ is a (graded) commutative algebra, with multiplications of vectors similar to the usual multiplication of two polynomials. Vectors in $H^{n}(X,\mathbb{Z}_2)$  are of order $n$, and the product $x_ny_{n'}\in H^{n+n'}(X,\mathbb{Z}_2)$ is of order $n+n'$ with $x_n\in H^n(X,\mathbb{Z}_2)$ and $y_{n'}\in H^{n'}(X,\mathbb{Z}_2)$. Hence, we can always expand a ``polynomial'' by increasing the orders. 

Consider a smooth parameter space $X$, over which $d_V$-dimensional ($d_V$D) real vector spaces $E_X(x)$ with $x\in X$ are continuously distributed. The $n$th SW class of $E_X$, denoted by $w_n(E_X)$, is an $n$D cohomology class, namely, $w_n(E_{X})\in H^{n}(X,\mathbb{Z}_2)$.
We may introduce the total SW class as a vector in $H^{*}(X,\mathbb{Z}_2)$,
\begin{equation*}
	w(E_X)=1+w_1(E_X)+w_2(E_X)+\cdots+w_d(E_X).
\end{equation*}
Here, $d=\mathrm{min}\{d_V,d_X\}$, because $H^{n}(X,\mathbb{Z}_2)=0$ if $n>d_X$, and $w_n(E_{X})$ vanishes if $n>d_V$.

\section{Numerical verification of topological asymmetry over the Klein bottle}\label{D}
In this section, we present a numerical method to verify the topological asymmetry condition $w_2^+ \neq w_2^-$ over the Klein bottle. Given that $w^\pm_1(c)=1$ prevents direct computation of $w_2^\pm$ via Wilson loop methods, we construct the Hamiltonian $\H_t(\k)$ defined as
\begin{equation}\label{eq:Ht}
	\H_t(k_x,k_y) = \begin{bmatrix}
		\H(k_x,k_y) & \Delta \\
		\Delta^\dagger & -\H(k_x+\pi,k_y)
	\end{bmatrix},
\end{equation}
where $\Delta$ denotes a perturbation satisfying the commutation relations $[PT,\Delta]=0$ and $[U_{M},\Delta]=0$. The nontrivial second Stiefel-Whitney number $w_2$ of $\H_t(\k)$ serves as an indicator for the topology asymmetry $w_2^+ \neq w_2^-$ in the original Hamiltonian $\H(\k)$.

To establish the connection between $\H_t(\k)$ and $\H(\k)$, we define the topological invariants of $\H(k_x,k_y)$ [$-\H(k_x+\pi,k_y)$] as $w_1^{\pm}(a/b/c)$ [$\nu_1^{\pm}(a/b/c)$] and $w_2^{\pm}$ [$\nu_2^\pm$]. These topological invariants satisfy the relations
\begin{equation}\label{eq:topology_relations}
	\begin{split}
		&w^\pm_1(c) = \nu^\pm_1(c) = 1, \quad w_2^\pm = \nu_2^\mp, \\
		&w^\pm_1(a) = \nu^\pm_1(b) = 1 - w^\pm_1(b) = 1 - \nu^\pm_1(a).
	\end{split}
\end{equation}
The total Stiefel-Whitney class $w^\pm = w(E^{\pm}_\mathcal{M}) \in H^*(X,\mathbb{Z}_2)$ for  conduction and valence bands of $\H_t(\k)$ take the form
\begin{equation}
	\begin{split}
		w^\pm &= (1 + w_1^\pm + w_2^\pm)(1 + \nu_1^\pm + \nu_2^\pm) \\
		&= 1 + w_1^\pm + \nu_1^\pm + w_1^\pm\nu_1^\pm + w_2^\pm + \nu_2^\pm.
	\end{split}
\end{equation}

Therefore, two crucial observations emerge from above formulation:
\begin{enumerate}
	\item The vanishing first SW number: $w^\pm_1(c)[\H_t(\k)] = w^\pm_1(c) + \nu_1^\pm(c) = 0$ enables Wilson loop computation of $w_2$.
	\item The product term vanishes: $w_1^\pm\nu_1^\pm = 0$ since $\alpha\beta=0$, where $\alpha$ and $\beta$ are the respective cocycles for $\H(k_x,k_y)$ and $-\H(k_x+\pi,k_y)$.
\end{enumerate}
The second SW number of $\H_t(\k)$ consequently reduces to
\begin{equation}
	w_2^\pm(\H_t) = w_2^\pm + \nu_2^\pm = w_2^\pm + w_2^\mp \mod 2.
\end{equation}
Finally, we obtain that
\begin{equation}
	w_2^\pm(\H_t) = \begin{cases}
		1 & \text{if } w_2^+ \neq w_2^- \\
		0 & \text{if } w_2^+ = w_2^-
	\end{cases}.
\end{equation}

Applying this framework to our theoretical model, we substitute the Hamiltonian $\H(\k)$ from Eq.~\eqref{eq:model} into Eq.~\eqref{eq:Ht} with the perturbation matrix,
\begin{equation}
	\Delta = 0.3\begin{bmatrix}
		0 & 0 & 1 \\
		1 & 0 & 0 \\
		1 & 0 & 0
	\end{bmatrix}.
\end{equation}
Figure~\ref{fige1}(d) displays the resulting Wilson loop spectrum, where the observed value $w_2(\H_t)=1$ provides numerical verification of the topological asymmetry in our original Hamiltonian $\H(\k)$ [Eq.~\eqref{eq:model}].

\section{The physical model for real Klein bottle insulator}\label{E}
	\begin{figure}[t]
	\includegraphics[width=\columnwidth]{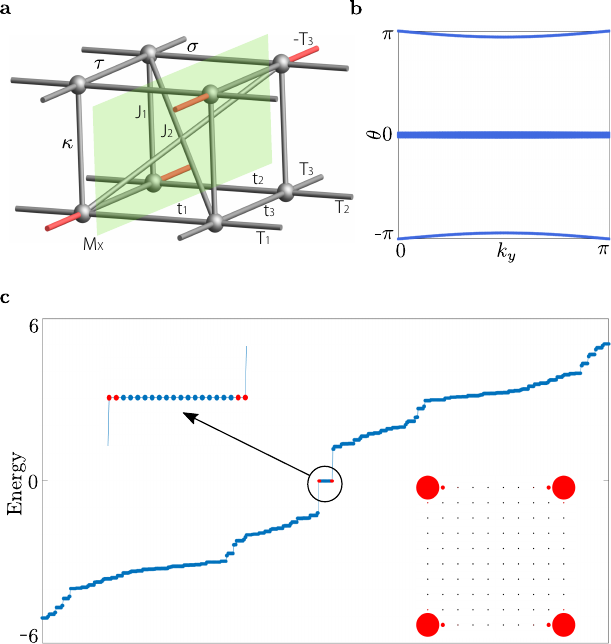}
	\caption{\textbf a Lattice model with $PT$ symmetry and mirror symmetry $\mathcal{M}_x$. \textbf b Wilson loop spectrum of $\H_t(\k)$ constructed from $\H(\k)$ in Eq.~\eqref{eq:H_phy}, where $w_2(\H_t)=1$ confirms the topological asymmetry $w_2^+ \neq w_2^-$.  \textbf c shows the calculated energy spectrum with opening the boundaries. Four corner states labeled by red dots are observed.}
	\label{fig:physicalmodel}
\end{figure}

The lattice model, depicted in Fig.~\ref{fig:physicalmodel}(a), exhibits inversion symmetry through the center of the unit cell. Each cell consists of eight sites, indexed by three qubits $\kappa$, $\tau$, and $\sigma$. The system possesses $PT$ symmetry with $(PT)^2 = 1$ and mirror symmetry $M_x$ that anticommutes with the translation $L_{y}$.

 The Hamiltonian in momentum space can be written as
\begin{equation}\label{eq:H_phy}
	\begin{split}
		\H(\k)&=f_1^{+}(k_x)\Gamma_{001}+f_1^{-}(k_x)\Gamma_{331}-f_2^{+}(k_x)\Gamma_{002}\\
		&-f_2^{-}(k_x)\Gamma_{332}+t_3\Gamma_{010}-T_3\cos k_y\Gamma_{313}\\
		&+T_3\sin 	k_y\Gamma_{323}+J_1\Gamma_{100}+\frac{J_2}{2}(\Gamma_{111}-\Gamma_{221}).
	\end{split}
\end{equation}
where $f_1^{\pm}(k_x)=(t_1\pm t_2)/{2}+(T_1\pm T_2)\cos k_x/{2}$, $f_2^{\pm}(k_x)=(T_1\pm T_2)/{2}\sin k_x$ and $\Gamma_{ijk}=\kappa_i\otimes\tau_j\otimes\sigma_k$. The symmetry operators are explicitly given by 
\begin{equation}
	PT=\Gamma_{111}\K,\quad M_x=\Gamma_{001}I_x\mathcal L_{k_y/2}.
\end{equation}

Taking the parameter values as
\begin{equation}
	(t_1,t_2,t_3,J_1,J_2,T_1,T_2,T_3)=(1,3,0,0,2,2,0,1),
\end{equation}
we numerically calculate the topological invariants and energy spectrums. The first SW numbers are $w_1(a)=1$, $w_1(b)=0$, and $w_1(c)=1$. To verify the topological asymmetry $w_2^+ \neq w_2^-$, we construct the Hamiltonian $\H_t(\k)$ as defined in Eq.~\eqref{eq:Ht}, using the perturbation matrix,
\begin{equation}
	\Delta = \sigma_1 \otimes \begin{bmatrix}
		1 & 1 & 1 & 1 \\
		1 & 1 & 1 & 1 \\
		1 & 1 & 1 & 1 \\
		1 & 1 & 1 & 1
	\end{bmatrix}.
\end{equation}
The Wilson loop spectrum of $\H_t(\k)$ (Fig.~\ref{fig:physicalmodel}(b)) demonstrates nontrivial topology with $w_2(\H_t) = 1$, confirming the predicted band asymmetry in $\H(\k)$.  As for the boundary states, we can observe four corner states labeled by the red dots in Fig.~\ref{fig:physicalmodel}(c). The other zero-mode states labeled by the blue dots are protected by the nontrivial $w_1$.
\color{black}

\newpage
\bibliography{references}

\end{document}